%% file: main.tex
\documentclass[sigplan,nonacm]{template/acmart}
\usepackage{xspace}
\usepackage{pifont}
\usepackage{booktabs}
\usepackage{multirow}
\usepackage{url}
\usepackage{soul}
\usepackage{comment}
\usepackage[font=footnotesize,labelfont=bf]{caption}
\usepackage{subcaption}
\usepackage{tcolorbox}
\tcbuselibrary{skins}

\newcommand{\asaf}[1]{\textcolor{blue}{\{Asaf: #1\}}}

\newcommand{\sherry}[1]{\textcolor{orange}{\{Sherry: #1\}}}

\usepackage{enumitem}
\setitemize{noitemsep,topsep=0pt,parsep=0pt,partopsep=0pt}
\newenvironment{denseitemize}{
\begin{itemize}[topsep=2pt, partopsep=0pt, leftmargin=1.5em]
  \setlength{\itemsep}{2pt}
  \setlength{\parskip}{0pt}
  \setlength{\parsep}{0pt}
}{\end{itemize}}

\definecolor{sherry}{RGB}{255,140,0}
\definecolor{dong}{RGB}{0,0,200}
\definecolor{checked}{RGB}{0,0,0}
\definecolor{check}{RGB}{255,0,0}

\newtcolorbox{observationbox}[1][]{%
  enhanced,
  colback=gray!10,
  colframe=gray!60,
  boxrule=0.8pt,
  arc=2pt,
  left=6pt, right=6pt, top=4pt, bottom=4pt,
  fonttitle=\bfseries,
  #1
}
\newcounter{observation}
\newcommand{\observation}[1]{%
  \refstepcounter{observation}%
  \begin{observationbox}%
  \textbf{Observation \theobservation:} #1%
  \end{observationbox}%
}

\renewcommand\footnotetextcopyrightpermission[1]{}

\AtBeginDocument{%
  }

\begin{document}

\newcommand{\name}{TierBPF\xspace}
\title{\name: Page Migration Admission Control for Tiered Memory via eBPF}

\author{Xi Wang}
\affiliation{%
  \institution{University of California, Merced}
  \country{USA}
}

\author{Tal Zussman}
\affiliation{%
  \institution{Columbia University}
   \country{USA}
}

\author{Bin Ma}
\affiliation{%
  \institution{University of California, Merced}
 \country{USA}
}

\author{Asaf Cidon}
\affiliation{%
  \institution{Columbia University}
\country{USA}
}

\author{Dong Li}
\affiliation{%
  \institution{University of California, Merced}
  \country{USA}
}
\renewcommand{\shortauthors}{Wang et al.}

\begin{abstract}
Software-based memory tiering systems decide which pages to place on the slower or faster tier. However, they do not  consider two important factors that greatly influence application performance: the size of the migrated pages, and the underlying hardware device and tiering topology. We introduce \name, a software mechanism that can be plugged into existing memory tiering systems to take
these factors into account,  
by making 
\textcolor{checked}{architecture-aware} 
page admission decisions. 
\name is implemented as a set of eBPF hooks, allowing users to define their custom policies.
In order to make its decisions, \name utilizes a lightweight tracking mechanism for page profiling which is not dependent on the application's working set size. \name, integrated into three memory tiering systems and evaluated with 17 workloads, achieves geomean throughput gains of up to 15.9\% with improvements of up to 75\% for individual workloads.
\end{abstract}

\maketitle
\pagestyle{plain}

\input{text/intro}
\input{text/background}

\input{text/motivation}

\input{text/overview}
\input{text/design}

\input{text/implementation}
\input{text/evaluation}
\input{text/related_work}

\input{text/conclusion}

\clearpage
\balance
\bibliographystyle{template/ACM-Reference-Format}
\bibliography{bib/sherry,bib/li}

\end{document}

%% file: text/intro.tex
\section{Introduction}
\label{sec:intro}


As the number of cores available in modern processors grows faster than local DRAM server capacity~\cite{maruf2023tpp,10.5555/3691938.3691941}, 
datacenter operators are deploying tiered-memory architectures that pair fast, capacity-limited local DRAM with a slower, higher-capacity memory tier. \emph{Memory tiering systems} are responsible for migrating pages across these tiers.

Memory tiering systems are typically designed around a high-level abstraction: 
a heterogeneous memory architecture composed of slow memory and fast memory. This abstraction simplifies the page migration decision based on page access frequency and recency~\cite{maruf2023tpp, vuppalapati2024colloid, xiang2024nomad,10.1145/3676642.3736119,eurosys24:mtm,atc24_hm}.   
That is, migrating hot (frequently or recently accessed) pages to the fast tier, and keeping cold (infrequently used) pages in the slow tier.  Albeit simple to reason about, this high-level abstraction causes two problems. 

\textbf{Problem 1: The abstraction ignores the interaction between page size and migration.} Many production systems enable Transparent Huge Pages (THP), allocating memory in 2\,MB pages to reduce the TLB miss overhead. However, page migration in memory tiering systems is intertwined with the page size. When the kernel migrates a 2\,MB THP, it must copy the entire page, even if only a small subset of its constituent base pages is frequently accessed. This behavior wastes migration bandwidth and may squander 
scarce fast-tier capacity with cold data. Splitting the THP into 512 base pages and migrating only the hot ones avoids this waste, but at the price of increased TLB overhead. 

The problem of granular hot memory regions in THP 
is not specific to migration. Indeed, Linux v6.8 attempted to address this by introducing multi-size THP (mTHP)~\cite{linux_mthp}, which supports intermediate page sizes from 16\,KB to 1\,MB. While easier to allocate under memory fragmentation than a 2\,MB page, an intermediate size retains two benefits of a huge page: 
(1) fewer page faults 
and (2) larger TLB reach, exemplified by  ARM64's contiguous-PTE hint~\cite{arm_arm} and AMD's PTE coalescing~\cite{amd_zen5_sog,solomon2023pte}. 
However, with mTHP, the page size is determined only at \emph{allocation} time. As such, no existing tiering system exploits splitting pages into intermediate 
sizes during page migration, and  
a page originally allocated as a 1\,MB page can only be migrated in one big chunk. 


\textbf{Problem 2: The fast-slow tier abstraction hides memory architectural properties.} Tiering deployments are heterogeneous, both in terms of the types of devices and in terms of the deployment topology. CXL-attached memory expansion connects over a PCIe link, Intel Optane PMEM is integrated via the DDR bus, while emerging substrates, such as CXL-attached flash~\cite{288786,10.1145/3669940.3707250} and disaggregated memory pools~\cite{10.1145/3575693.3578835,10.1145/3722212.3724460,10.1145/3712285.3759816}, further broaden the design space. These deployment configurations differ not only in latency and bandwidth, but also in architectural properties such as memory duplex mode (i.e., read-write interference), persistence, and cache coherence. As a result, there is no one-size-fits-all policy that works for any deployment configuration. For example, a policy that improves performance on a CXL-based memory expansion may have little benefit---or even incur overhead---on persistent memory-based platforms (\S\ref{subsec:mot-arch}). 

These two problems share a common root cause: there is no existing mechanism for tiered memory systems to take page size or the tiered memory's deployment configuration into account. 
To this end, we introduce \name, a system that interfaces with existing tiered memory systems, making them adaptable to factors like page size granularity and tiering architecture. \name sits between the memory tiering system's migration decision-making and the page migration mechanism. Our \textbf{key insight} is that \name makes 
existing tiered memory systems aware of their broader context by applying \emph{page migration admission control}. When the tiering system decides that a page should be migrated, \name applies two \emph{filters} before the migration proceeds: a \emph{granularity filter} that determines the right page size for migration, and an \emph{architecture filter} that determines whether the migration actually benefits the deployed hardware. These two filters transform a coarse migrate-or-not decision into a fine-grained one that accounts for subpage access patterns, runtime conditions (i.e., memory bandwidth contention), and hardware characteristics (i.e., memory duplex mode). 

{The granularity filter addresses Problem~1.} Instead of migrating the full 2\,MB THP or splitting it all the way to 4\,KB, this filter selects an intermediate mTHP size large enough to 
reduce TLB and page-fault costs yet small enough to avoid migrating cold data. The choice is \emph{contention-aware}: under low slow-tier bandwidth contention, the filter keeps pages at 2\,MB for 
minimum TLB and page-fault costs; under high contention---common in multi-tenant cloud environments---
the filter selects the largest mTHP size that tightly captures a THP's hot region (e.g., 64\,KB or 512\,KB) (\S\ref{subsec:mot-contention}).


The granularity filter yields two performance benefits. First, memory is allocated at 2\,MB granularity by default, and a THP is split only when the distribution of its hot regions is sparse enough. Many THPs therefore remain 2\,MB-mapped, retaining TLB coverage and low page-fault overhead (\S\ref{sec:eval_tlb_fault}). Second, migrating only a THP's hot region rather than the entire 2\,MB THP improves migration efficiency: migration moves less data, wastes less bandwidth, and leaves more fast-tier capacity for hot data (\S\ref{sec:eval_deep}).

The architecture filter addresses Problem~2. The specific device hardware affects the effectiveness of page migration. For example, CXL memory uses full-duplex links where reads and writes proceed on independent channels~\cite{yang2025cxlaimpodcxlmemoryneed}, while Intel Optane PMEM uses half-duplex DDR where reads and writes share the same bus. Hence, on a CXL system with read-dominant memory traffic, selectively holding back write-heavy pages from page promotion can improve bandwidth utilization by keeping the full-duplex channels balanced. The same policy on PMEM would merely prevent hot pages from reaching fast memory, with no bandwidth benefit. The page migration admission control allows memory tiering to customize migration decisions based on the duplex mode.

A critical mechanism in supporting these filters is accurate and low-overhead memory profiling. Building and maintaining per-page profiling 
(as in MEMTIS~\cite{lee2023memtis}) 
cannot support the decisions for smaller page sizes because of sampling sparsity during profiling (\S\ref{subsec:profiling}) and profiling  metadata growth with the working set. 
Instead, \name introduces a lightweight, compact global subpage histogram (using only 4\,KB per process), 
enabling lightweight profiling with overhead independent of the workload's working set size. In addition, \name uses eBPF-based profiling, processing hardware memory-access samples in-kernel and avoiding context switches and data copying suffered by user-space profiling. 

To make page-migration admission control deployable across different tiering systems and adaptable to evolving memory architectures, we implement all policy logic as user-customizable eBPF programs. 
We expose three hooks along the kernel's NUMA migration path: one for split-size selection, one for subpage hot/cold classification, and one for migration admission. 
This design cleanly separates policy (i.e., which page to migrate and at what granularity) from mechanism (i.e., how to migrate), allowing the admission control to easily plug into existing memory tiering systems.

Our paper's major contributions are:


\begin{denseitemize}
    \item We identify two problems in existing memory tiering systems caused by the fast-vs.-slow memory abstraction: the lack of both migration-time page-granularity awareness and architecture-aware page migration filtering. 

    \item We present \name, the first system to support dynamic, contention-aware mTHP-size selection and memory duplex mode-aware page migration. \name implements all migration policies as eBPF programs, allowing operators to adapt migration policies to hardware topologies.


    \item We evaluate \name on two memory platforms: (a) CXL memory expansion, and (b) Intel PMEM. We integrate \name with three memory tiering systems: AutoNUMA~\cite{autonuma}, TPP~\cite{maruf2023tpp}, and Colloid~\cite{vuppalapati2024colloid}. Our results show that \name significantly improves both adaptability and performance across architectures, outperforming prior approaches under diverse workloads: \name achi-eves geomean throughput gains of 9.3--15.9\%, with improvements of up to 75\% for individual workloads.    


\end{denseitemize}


%% file: text/background.tex
\section{Background}
\label{sec:background}

\textbf{Multi-size transparent huge pages.} Virtual-to-physical address translation depends on the TLB. 
Using 4\,KB pages limits TLB reach (e.g., a few hundred kilobytes per TLB), increasing TLB miss rates for memory-intensive workloads. Since each TLB miss invokes a multi-level page-table walk that may cost tens to hundreds of cycles depending on cache residency, address-translation overhead can reduce application performance by up to 30\%~\cite{weisberg2008obsolete,lustig2013tlb,gandhi2014badgertrap}.
Transparent huge pages (THPs) alleviate this by using 2\,MB pages, reducing translation frequency and improving TLB hit rates. 


However, 2\,MB THPs introduce their own challenges. Allocating physically contiguous 2\,MB regions becomes costly when memory is fragmented. To address this, Linux 6.8 introduced mTHP, which supports anonymous page allocation at intermediate sizes (16\,KB, 32\,KB, 64\,KB, 128\,KB, 256\,KB, 512\,KB, and 1\,MB) alongside the traditional 4\,KB and 2\,MB options. Each size can be enabled or disabled independently via sysfs. 
mTHP provides a practical middle ground: smaller huge-page sizes are easier to allocate, reduce memory waste, and still trigger fewer page faults and improve TLB efficiency compared to 4\,KB pages ~\cite{arm_arm,amd_zen5_sog,solomon2023pte}. 

The TLB implementation for intermediate sizes is architecture dependent. On ARM64~\cite{arm_arm}, the kernel marks an aligned, physically-contiguous PTE with a \emph{contiguous} hint that the MMU may cache as a single TLB entry. AMD processors since the Zen microarchitecture~\cite{amd_zen5_sog,solomon2023pte} instead perform PTE coalescing microarchitecturally, opportunistically storing aligned, physically-contiguous pages in a single TLB entry. Intel processors~\cite{intel_sdm} provide neither mechanism, extending TLB reach only at the 2\,MB PMD level. Internally, the Linux kernel represents each (m)THP as a \emph{folio}; we use folio and page interchangeably in this paper.

\begin{figure}[!t]
  \centering
    \includegraphics[width=1\linewidth]{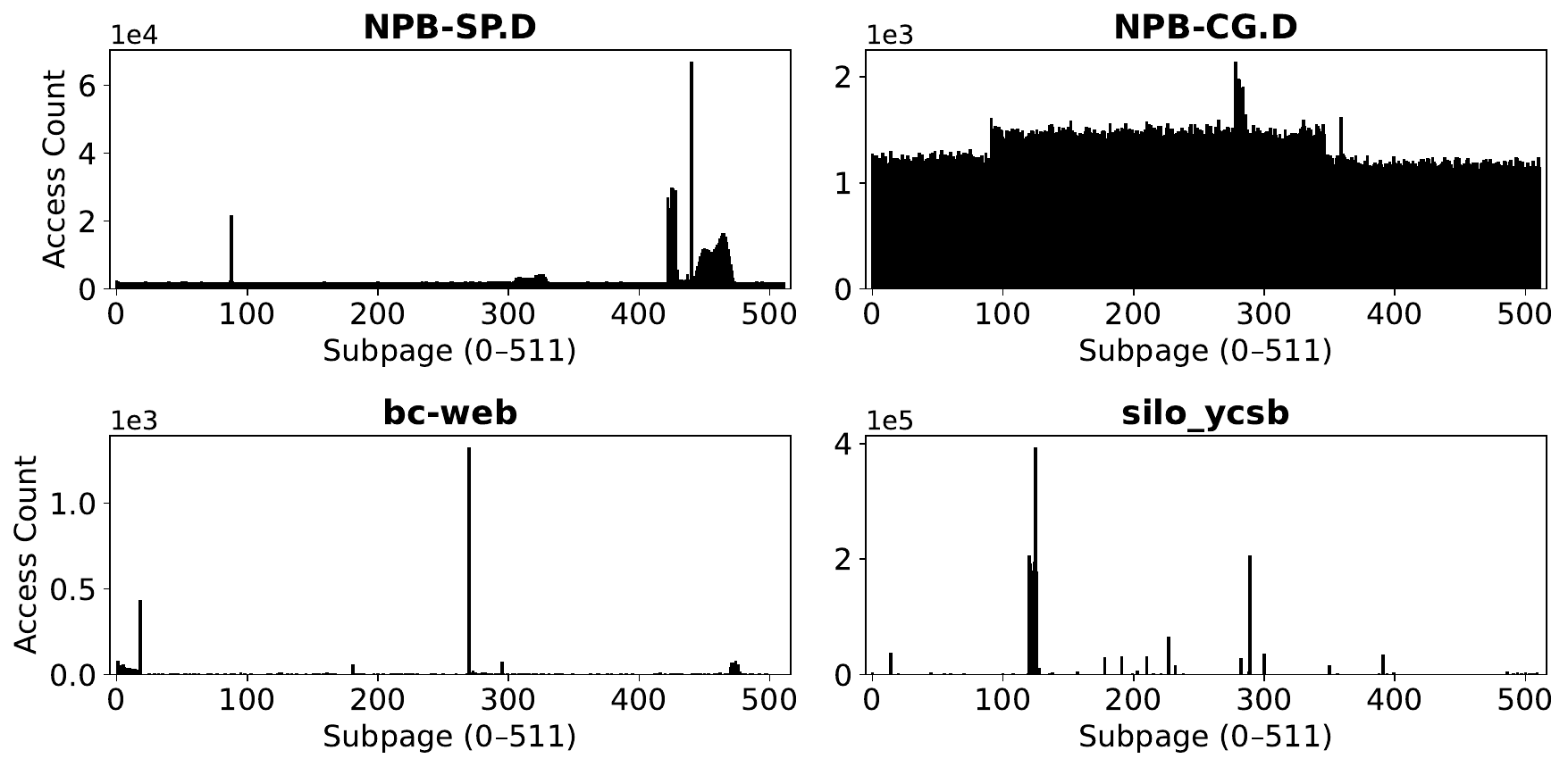}
\vspace{-20pt}
  \caption{Hot subpage distribution within 2\,MB THPs. We use workloads summarized in Table~\ref{tab:benchmarks}.}
  \label{fig:hot-subpage-dist}
  \vspace{-10pt}
\end{figure}

However, current mTHP support is limited to \emph{allocation}. The kernel does not consider 
page sizes when \emph{migrating} pages across memory tiers. 
MEMTIS~\cite{lee2023memtis} compares the measured fast-tier hit ratio against an estimated base-page (4\,KB page) hit ratio to make a \emph{binary} choice:
migrating the entire 2\,MB page or splitting it into 512 individual 4\,KB base pages for selective migration, with no consideration of other mTHP sizes. However, 
splitting a 2\,MB page all the way down to 4\,KB pages 
loses the performance benefits in TLB coverage and page-fault reduction entirely, even when a 64\,KB or 128\,KB split would have been sufficient. 
This motivates supporting multi-size THP during page migration.

\textbf{eBPF for kernel customization.} Extended Berkeley Packet Filter (eBPF) 
enables user-defined programs to run safely inside the kernel in a sandboxed environment, without requiring kernel source modifications or custom kernel modules. The eBPF verifier enforces strict safety guarantees: programs cannot crash the kernel, access arbitrary memory, or execute unbounded loops. Originally designed for network packet filtering, eBPF has evolved into a general-purpose mechanism for extending kernel functionality~\cite{sched_ext,zussman2025cache_ext,mores2024ebpf} and making kernel memory management programmable.

%% file: text/motivation.tex
\section{Motivation}
\label{sec:motivation}




\subsection{THP Migration Dilemma}
\label{subsec:mot-dilemma}

When THP is enabled, the kernel manages memory in 2\,MB pages by default. 
Consider a 2\,MB THP where only a small set of subpages is frequently accessed.
Existing memory tiering systems face two suboptimal choices. 

\textbf{Option 1 (Linux default):} migrate the full 2\,MB THP. This wastes migration bandwidth
and consumes scarce fast-tier capacity by allocating 2\,MB of DRAM for cold data.  Also, 
migrating a 2\,MB page requires a contiguous 2\,MB free region on the destination node, which is often unavailable under memory fragmentation, causing migrations to fail (\S\ref{sec:eval_deep}).

\textbf{Option 2 (e.g., MEMTIS~\cite{lee2023memtis}):} split the THP into 4\,KB base pages and migrate the hot ones. While this preserves bandwidth and fast-tier space, 
it reduces the benefits of (intermediate) huge pages 
(\S\ref{sec:background}). 
After splitting, 
the increase in faults and TLB misses can offset the gains of page migration.



Figure~\ref{fig:hot-subpage-dist} shows the distribution of hot 4\,KB subpages
within 2\,MB THPs. Access is rarely uniform:  
most THPs have a concentrated hot region spanning tens to hundreds of kilobytes---far larger than a single 4\,KB page but far smaller than the full 2\,MB. This suggests that intermediate mTHP sizes (e.g., 64\,KB or 256\,KB) could precisely capture the hot region, migrating only what is necessary while preserving TLB coverage and page-fault benefits for the migrated region.
\vspace{-5pt}
\observation{Only a small subset of subpages within each THP is hot. Migrating at 2\,MB wastes bandwidth and fast-tier capacity; splitting to 4\,KB discards the benefits in TLB and page faults. Intermediate mTHP sizes can capture the hot region without either penalty.}
\label{obs:subpage-skew}

\subsection{There is No One Ideal Page Size}
\label{subsec:mot-contention}

Deciding the mTHP size is challenging. With different levels of memory contention, the same application prefers different page sizes to improve performance.

\begin{figure}[t]
  \centering
    \includegraphics[width=1\linewidth]{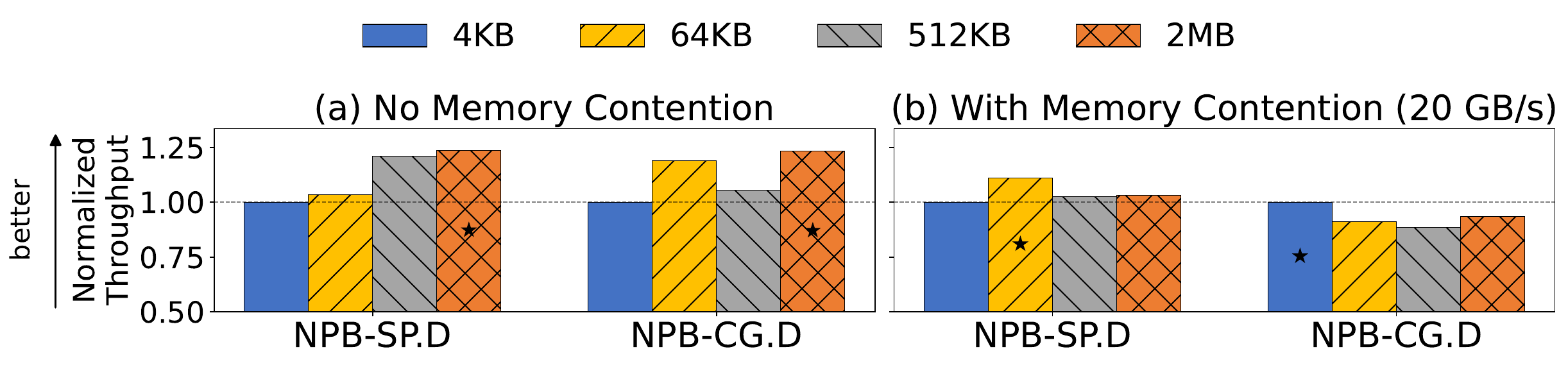}
    \vspace{-20pt}
  \caption{The optimal mTHP size shifts under memory contention, as per \S\ref{sec:eval_method}. \textcolor{checked}{Stars indicate the best performance.}}
  \label{fig:contention-size}
  \vspace{-15pt}
\end{figure}


When an application runs alone and the slow tier is not under bandwidth pressure, larger THP sizes are generally preferable. In this setting, the slow tier can offer relatively low memory-access latency~\cite{vuppalapati2024colloid}, 
so the cost of migrating huge pages is small, and the bandwidth consumed by such migrations does not meaningfully affect performance. As a result, the TLB benefits of 2\,MB pages dominate. However, in cloud environments where multiple workloads share the same machine and memory tiers, the trade-offs shift. Under contention in the slow tier, migrating 2\,MB pages becomes more expensive: misidentified hot pages waste migration bandwidth, and truly hot pages left in the slow tier suffer greater performance penalties due to increased access latency. In these scenarios, smaller mTHP sizes reduce both slow-tier bandwidth consumption and fast-tier capacity usage per migration, allowing more distinct hot regions to reside in fast memory simultaneously.


Figure~\ref{fig:contention-size} shows the optimal mTHP size under memory contention. The results show that there is no one-size-fits-all size: the optimal page size is different when varying the workloads and memory contention level. This implies that any static page-size policy---whether configured system-wide via sysfs or defined per application---will be suboptimal in a dynamic, multi-tenant environment. The system must instead adapt its mTHP size selection at runtime, guided by the memory architecture and prevailing memory pressure.

Linux mTHP~\cite{linux_mthp} allows administrators to enable or disable specific mTHP sizes via sysfs, but this is a static, manual configuration. It does not adapt to runtime conditions. eBPF-mm~\cite{mores2024ebpf} uses eBPF to select page sizes at allocation time, but it relies on pre-defined application profiles and does not consider page migration. 
Neither system provides \emph{dynamic}, contention-aware mTHP size selection for page migration.

\vspace{-5pt}
\observation{The optimal mTHP size depends on workload and runtime memory contention.
Under low contention, larger pages are preferred for higher TLB coverage and fewer page faults; 
under high contention, smaller mTHP sizes cause less overhead, waste less fast-tier capacity, and yield better overall performance.
A static size policy cannot adapt to these changes.}
\label{obs:contention}

\subsection{Architecture-Agnostic Migration}
\label{subsec:mot-arch}

\begin{figure}[!t]
  \centering
    \includegraphics[width=1\linewidth]{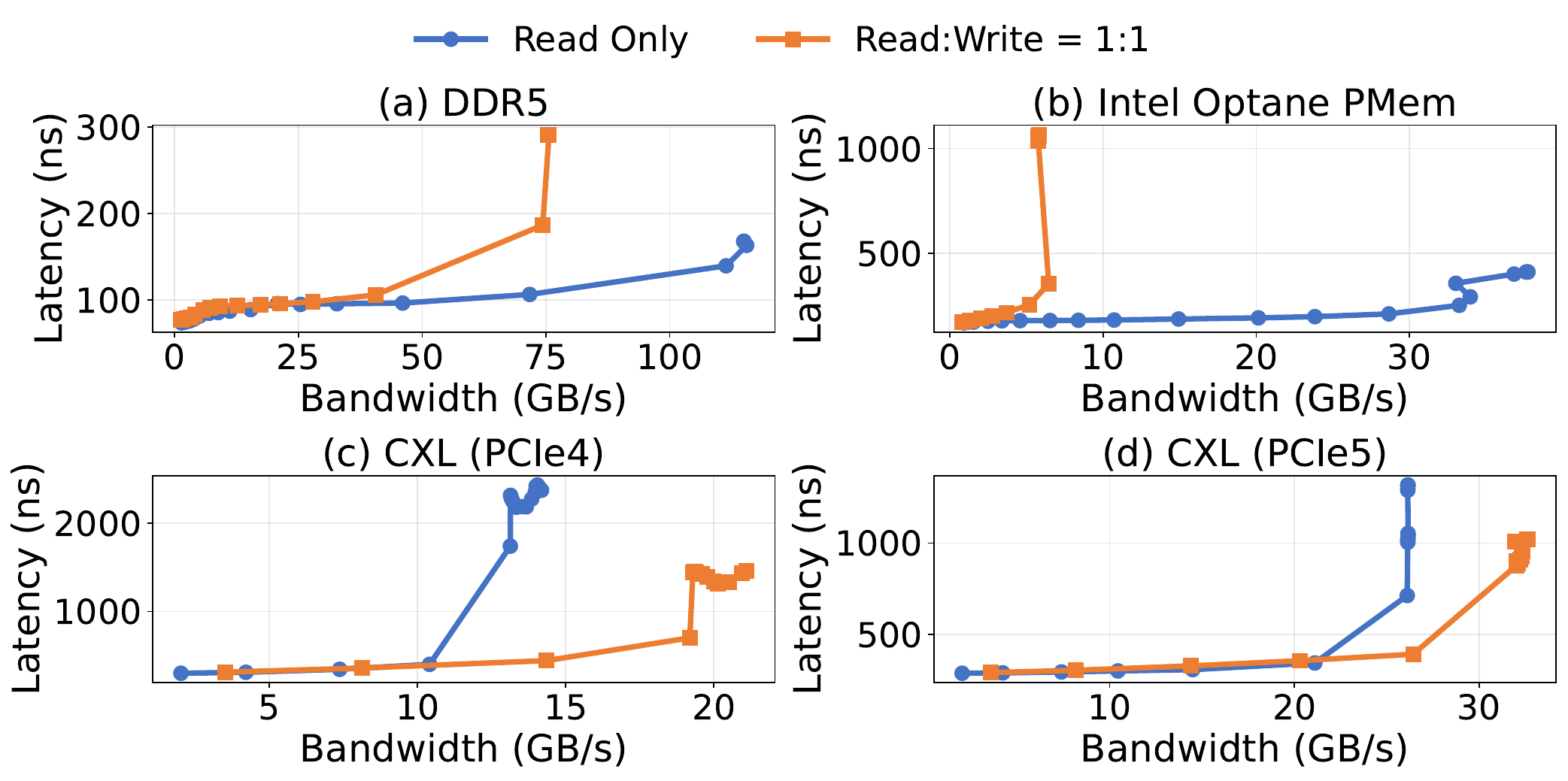}
  \vspace{-20pt}

  \caption{Bandwidth--latency varies with the R/W mix.}
  \label{fig:arch-bandwidth}
  \vspace{-15pt}
\end{figure}

Existing tiered memory systems apply the same migration policy without considering the underlying memory architecture. 
We use CXL memory and Intel Optane PMEM as a case study.
CXL memory provides full-duplex links where reads and writes use independent channels~\cite{yang2025cxlaimpodcxlmemoryneed}. 
The migration policy can take advantage of this property by rebalancing the read/write traffic across the full-duplex channels. 
PMEM has half-duplex DDR interfaces where read and write traffics compete for the same link. As a result, rebalancing read/write traffic does not improve bandwidth utilization but shifts contention from one direction to the other. Worse, selectively holding back read- or write-heavy pages from promotion prevents genuinely hot pages from reaching the fast tier, degrading overall performance.  

As a result, a traffic‑rebalancing policy that performs well on a CXL platform can lose effectiveness on a PMEM platform. Memory‑tiering systems must therefore be tailored to the memory architecture. Figure~\ref{fig:arch-bandwidth} illustrates this discrepancy: CXL \textcolor{checked}{reaches 149\% and 125\% of the read-only peak bandwidth on PCIe4 and PCIe5, respectively,} with the mixed read/write pattern by leveraging its full‑duplex link, whereas PMEM, constrained by its half‑duplex DDR interface, experiences substantial bandwidth degradation (\textcolor{checked}{only 17\% of the read-only peak bandwidth}) under the same traffic pattern.


\observation{Migration policies produce different outcomes on different hardware configurations. We must customize the policy to specific tiered-memory hardware setups to maximize performance.} 
\label{obs:arch}

%% file: text/overview.tex
\section{Key Ideas in \name}
\label{sec:key-ideas}

\name addresses the three gaps identified in \S\ref{sec:motivation}. 
(1) It splits 2\,MB THPs into optimal mTHP sizes based on hot subpage distribution 
(\S\ref{subsec:mot-dilemma}). 
(2) It activates THP splitting based on runtime memory contention (\S\ref{subsec:mot-contention}).   
(3) It customizes migration decisions based on memory architecture characteristics (\S\ref{subsec:mot-arch}). The three policies are eBPF programs, 
making them safe, low-overhead, and deployable without further kernel modifications.

\textbf{Lightweight memory profiling.} 
Every policy decision in \name---which mTHP size to split to, which subfolios to promote, whether to admit a migration---relies on access profiles collected from hardware event sampling. The profiling must be both effective and cheap.
\name maintains a \emph{global subpage histogram}---a fixed 512-entry array (4\,KB) per application---that \textcolor{checked}{maps} hardware-sampled accesses from all THPs \textcolor{checked}{to} their subpage offset. This constant-sized structure does not grow with the working set of the workload. The profiling logic runs as an eBPF program attached to hardware performance events, aggregating samples directly in interrupt context and avoiding context switches and data copies between kernel and user space (\S\ref{subsec:profiling}). 

\textbf{Contention-aware mTHP splitting.} \name dynamically selects the \emph{largest} mTHP size that captures most of the hot subpages within a dense memory region. The mTHP size is not static. 
Because the optimal size changes as memory contention varies (Observation~\ref{obs:contention}), \name activates 
THP splitting only when the slow-tier bandwidth is under pressure. 
\textcolor{checked}{Beyond reducing bandwidth waste, splitting also reduces migration failures: a smaller subfolio requires far less contiguous free space than a 2\,MB THP, so migrations succeed more often under memory fragmentation.}


\textbf{Architecture-aware migration policy.}  Even when a page is hot enough to merit promotion, migrating it does not always help. On a CXL system with full‑duplex channels, reads and writes proceed independently. If CXL read bandwidth is the bottleneck, promoting read‑heavy pages alleviates pressure, but promoting write‑heavy pages offers little benefit, because the write channel is typically uncongested (Observation~\ref{obs:arch}). On a half‑duplex PMEM system, however, the same selective policy would be counterproductive: reads and writes share a single channel, so all hot pages benefit equally from promotion.


\name adds an architecture‑aware migration‑admission hook. Before a page is promoted, an eBPF program evaluates its read/write ratio in the context of real‑time CXL traffic. Write‑heavy pages are deferred unless the CXL write channel is genuinely saturated. This hook is enabled only on CXL platforms and becomes a no‑op on PMEM (or other half-duplex memory architectures), or single‑tier systems.



\input{table/ebpf_hooks}

\begin{figure}[!t]
  \centering
  \includegraphics[width=0.9\columnwidth]{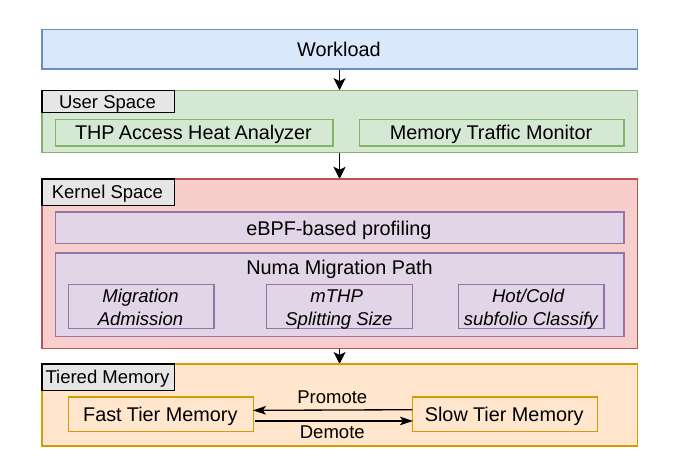}
  \vspace{-20pt}
  \caption{\name architecture overview.}
  \label{fig:overview}
  \vspace{-10pt}
\end{figure}

\textbf{eBPF as the policy layer.} \name uses eBPF to decouple migration policy from  migration mechanism. It exposes three hooks (Table~\ref{tab:hooks}) along the 
NUMA migration path as eBPF attachment points: one for split‑size selection, one for subpage hot/cold classification, and one for migration admission. When no eBPF program is loaded, the kernel falls back to safe defaults (no splitting and unconditional migration). 


The architecture of \name spans three layers (Figure~\ref{fig:overview}).
In the kernel, an eBPF program attached to hardware performance events profiles memory accesses and aggregates them into BPF maps. 
User‑space daemons derive policy decisions from the profiling data and publish them to the pinned BPF maps. On the kernel's NUMA migration path, three eBPF hooks serving as page migration admission control enforce these decisions at migration time. 


%% file: table/ebpf_hooks.tex
\begin{table}[!t]
\centering
\footnotesize
\caption{Three eBPF hooks in \name.}
\vspace{-10pt}
\label{tab:hooks}
\setlength{\tabcolsep}{2pt}
\begin{tabular}{lll}
\toprule
\textbf{Hook} & \textbf{Location} & \textbf{Default} \\
\midrule
\texttt{bpf\_pick\_split\_order} & \texttt{mm/migrate.c} & 2MB (no split) \\
\addlinespace
\texttt{bpf\_subpage\_is\_cold} & \texttt{mm/migrate.c} & 0 (all hot) \\
\addlinespace
\texttt{bpf\_numa\_migrate\_admission} & \texttt{sched/fair.c} & 0 (allow) \\
\bottomrule
\end{tabular}
\vspace{-1.5em}
\end{table}

%% file: text/design.tex
\section{\name Design}
\label{sec:design}

\subsection{Lightweight Memory Profiling}
\label{subsec:profiling}


\subsubsection{Global Subpage Histogram in mTHP Splitting}
\label{subsubsec:histogram}

For mTHP splitting, what matters is not the exact access count of each individual page, but the \emph{access distribution pattern within a 2\,MB THP region} (i.e., which subpages are hot/cold). Instead of creating a subpage histogram of 512 entries per huge page, \name captures hot subpage distribution with a global subpage histogram: an array of 512 entries (representing 512 subpages) per application, where each hardware event sample is mapped to a slot by \texttt{page\_addr \% 512}, 
\textcolor{checked}{which maps} accesses from all 2\,MB THPs of an application into one 512-slot array, requiring only $512 \times 8\,\text{B} = 4\,\text{KB}$ per application regardless of working set size.

\begin{figure}[t]
  \centering
  \includegraphics[width=1\linewidth]{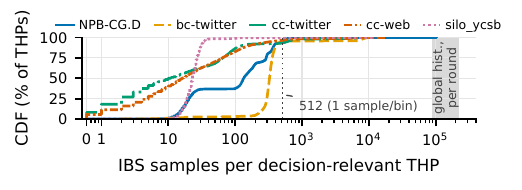}
  \vspace{-25pt}
  \caption{CDF of hardware samples accumulated per THP for decision-relevant (NUMA-faulted) THPs. The dashed line marks 512 samples (one sample per histogram bin on average). The shaded band marks the sample mass ($0.9\times10^5$--$2.1\times10^{5}$) the global histogram holds per decision round.}
  \label{fig:perthp-sparsity-cdf}
  \vspace{-1em}
\end{figure}

\textbf{Why not a per-THP histogram?}
The memory overhead of the per-THP histogram grows with the workload's working set size, which is not scalable. Even worse, the per-THP histogram faces a \emph{sampling sparsity} problem. Because the hardware sampling budget is spread across every THP of the application, any single THP accumulates far too few samples to populate a 512-bin distribution. Figure~\ref{fig:perthp-sparsity-cdf} shows this fact across five workloads: even when samples are accumulated over the \emph{entire run}, 93--100\% of decision-relevant THPs receive fewer than 512 samples, i.e., less than one sample per bin. The per-THP histogram is starved further by recency decay: replaying the production decay on per-THP counters leaves a median of only 1--9 samples per THP at decision time. This sampling sparsity precludes reliable splitting decisions.

\textbf{Why can a global histogram work?} Folding all THPs into one histogram loses access distribution patterns for individual THPs.  
However, the key insight is that 
with sampling-based profiling, the hardware sampler naturally focuses on frequently accessed pages. The global histogram is therefore dominated by the subpage distribution of \emph{hot} THPs---precisely the ones that are candidates for migration. Cold THPs contribute few samples and have little influence on the histogram. Since cold THPs should remain in slow memory and are not to be migrated, their subpage distribution is irrelevant to splitting decisions.
In other words, the global histogram is not a low-quality approximation to access distribution patterns. Rather, it is a direct representation of the THPs that matter for migration. We validate the fidelity of the global histogram in \S\ref{sec:eval_fidelity}.

\subsubsection{eBPF-Based Kernel Profiling}
\label{subsubsec:ebpf-profiling}
Prior tiering systems~\cite{10.1145/3676642.3736119,raybuck2021hemem} process hardware samples at user space: the kernel serializes each sample into a perf ring buffer, and a user-space thread polls the buffer and parses each sample before updating its metadata, incurring context switches, data copies, and cache pollution along this per-sample path. \name instead runs the profiling logic in the kernel as an eBPF program attached to perf events: when a sampling interrupt fires, the program runs in interrupt context, reads the sampled address, and updates the BPF map, so raw samples never leave the kernel. User space reads the BPF maps only when it makes a policy decision, and it reads aggregated entries (hundreds) rather than raw samples (potentially millions).

\subsubsection{R/W Classification in Migration Admission}
\label{subsubsec:rw-class}

\textcolor{checked}{For duplex-aware migration admission (\S\ref{subsec:admission}), \name must classify each page's read/write ratio. This is inherently a \emph{per-page} query that the global histogram cannot serve. Explicitly maintaining per-page counters is not scalable (\S\ref{sec:background}). 
\textcolor{checked}{To address this problem, \name introduces a \emph{blocked counting Bloom filter} (bCBF) inspired by HybridTier~\cite{10.1145/3676642.3736119}. A counting Bloom filter (CBF) provides space-efficient approximate count estimation; a blocked CBF further improves cache efficiency by partitioning counters into cache-line-sized blocks, so each lookup touches exactly one cache line.} Concretely, a bCBF maps each page address via hashing to $k$ counters within a block. A \textsc{Get} returns the minimum of the $k$ counters; an \textsc{Update} increments only the counters at the current minimum~\cite{10.1145/3676642.3736119}.}

\textcolor{checked}{\name extends this mechanism with a \emph{dual}-bCBF design, one for reads and one for writes. The write ratio of a page is computed as $w_{\min} / (w_{\min} + r_{\min})$, where $w_{\min}$ and $r_{\min}$ are the \textsc{Get} results from the write and read bCBFs, respectively.}

\name allocates the dual bCBF only when CXL memory is present and duplex-aware admission is active.
On PMEM or single-tier systems, \name does not need this counting mechanism, \textcolor{checked}{because the half-duplex interface does not benefit from read/write traffic rebalancing (\S\ref{subsec:mot-arch}).}

\subsection{mTHP Splitting}
\label{subsec:splitting}

Given the subpage histogram from \S\ref{subsec:profiling}, \name must decide what mTHP size to split into, which subfolios are hot, and which subfolios to migrate. \textcolor{checked}{We do not change the Linux kernel's inherent page merging mechanism.}

\subsubsection{THP Splitting Decision}
\label{subsubsec:split-size}

When slow-tier bandwidth is contended and \textcolor{checked}{THP splitting is activated}, a background user-space analyzer reads the global subpage histogram for each application and computes the target mTHP size. Let $c_i$ denote the sampled access count of the $i$-th 4\,KB subpage within a 2\,MB THP region ($i = 0, 1, \ldots, 511$), and let $C_{\text{total}} = \sum_{i=0}^{511} c_i$.
The algorithm to split THP has four stages.

\textbf{Stage 1: hot threshold determination.}
This stage separates hot subpages from cold ones. \name sorts subpages by their access counts 
($\tilde{c}_0 \geq \tilde{c}_1 \geq \cdots \geq \tilde{c}_{511}$) and finds the smallest $K^*$ so that the cumulative sum of top-$K^*$ subpages reaches a target coverage ratio $P$ (default 80\%):
\vspace{-5pt}

\begin{equation}
\small
\sum_{j=0}^{K^*} \tilde{c}_j \;\geq\; P \cdot C_{\text{total}}
\label{eq:hot-threshold}
\end{equation}
\normalsize

The hot threshold is then $T = \tilde{c}_{K^*}$.
This coverage-based threshold adapts automatically to workload skew and varying access distributions, ensuring that the most frequently accessed subpages are consistently classified as hot.

\textbf{Stage 2: hot subfolio classification.}
Each subpage is classified using the threshold $T$:
\begin{equation}
\small
h_i =
\begin{cases}
1 \;(\text{hot}), & \text{if } c_i \geq T \\
0 \;(\text{cold}), & \text{if } c_i < T
\end{cases}
\label{eq:hot-cold}
\end{equation}
\normalsize
producing a binary hot/cold map for 512 subpages in a THP.

\textbf{Stage 3: mTHP size selection via heat density.}
The goal of this stage is to select the largest mTHP size that still tightly captures a hot region. For each candidate size (2\,MB, 1\,MB, 512\,KB, 256\,KB, 128\,KB, 64\,KB), the THP is partitioned into aligned subfolios of $L$ contiguous subpages, where $L$ equals the candidate size divided by 4\,KB (e.g., $L = 128$ for a 512\,KB subfolio). The heat density of a subfolio is defined as the fraction of its subpages classified as hot:
\begin{equation}
\small
    \text{heat\_density}([i,\, i{+}L)) = \sum_{j=i}^{i+L-1} h_j \;/\; L
\end{equation}
\normalsize

\name selects the largest mTHP size for which at least one aligned subfolio achieves a heat density of at least $\tau_h$. \textcolor{checked}{A lower $\tau_h$ would admit subfolios with excessive cold data, wasting fast-tier capacity; a higher $\tau_h$ would force unnecessarily small splits, sacrificing TLB coverage. By default $\tau_h = 75\%$, keeping  selected subfolios predominantly hot while still favoring larger mTHP sizes for less TLB pressure.} 

\textbf{Stage 4: migration target selection.}
After splitting, \name does not migrate all subfolios. A subfolio is added to the migration list if it satisfies either of two conditions: (1) it is classified as hot based on heat density, or  
(2) it contains the NUMA‑faulted address that triggered the migration.  The NUMA fault reflects a recent locality mismatch that the hardware event sampler may not yet have observed. By combining these two signals, \name selects subfolios using both frequency-based information (from the subpage histogram) and recency-based information (from the NUMA fault). Subfolios that are cold and do not contain the faulting address remain in slow memory.

\textbf{Flat distribution detection.}
Before executing the stages above, \name first checks whether the access distribution is nearly uniform using normalized Shannon entropy ($H_{\text{norm}}$) ~\cite{pielou1966measurement}, \textcolor{checked}{an information-theoretic measure that quantifies the spread of a probability distribution as a single scalar in $[0, 1]$. Applied here, it captures how evenly accesses are distributed across the 512 subpages. 0 means all accesses hit one subpage, 1 means perfectly uniform}:
\begin{equation}
\small
H_{\text{norm}} = \bigl({-\sum_{i} p_i \log_2 p_i}\bigr) \;/\; {\log_2 512}, \quad p_i = c_i / C_{\text{total}}
\label{eq:entropy}
\end{equation}
If $H_{\text{norm}} \geq 0.95$, the distribution is essentially uniform---there is no hot region to isolate, so splitting provides no benefit.
In this case, \name falls back to the default NUMA fault behavior: the entire THP is migrated without splitting, following the recency-based signal from the fault.


\begin{table}[!t]
\centering
\small
\caption{Memory duplex mode-aware migration admission policy on the CXL memory. R=read and W=write.}
\label{tab:admission-policy}
\begin{tabular}{lccc}
\toprule
 & \textbf{R-dominant} & \textbf{Balanced} & \textbf{W-dominant} \\
 & \textbf{CXL traffic} & \textbf{CXL traffic} & \textbf{CXL traffic} \\
\midrule
R-heavy page & Allow & Allow & Prevent \\
W-heavy page & Prevent & Allow & Allow \\
\bottomrule
\end{tabular}
\end{table}

\subsubsection{Kernel-Side Enforcement}
\label{subsubsec:split-migrate}

When a NUMA fault triggers migration of a THP and splitting is enabled, the kernel calls the first eBPF hook (\texttt{bpf\_pick\_split\_order}) to obtain the target mTHP splitting size.
It then splits the THP into subfolios of that order and calls the second hook (\texttt{bpf\_subpage\_is\_cold}) for each subfolio to detect hot subfolios.
Hot subfolios and the faulting subfolio are added to the migration list; cold subfolios are left in place.

\subsection{Duplex-Aware Migration Admission Control}
\label{subsec:admission}
The splitting pipeline above determines how to migrate (sub)pages. 
We also determine \emph{whether} to migrate or not. 

A user-space daemon monitors the CXL read/write traffic by hardware event counters and stores a sliding window of measurements in a BPF map. On each NUMA fault, the admission eBPF hook (\texttt{bpf\_numa\_migrate\_admission}) classifies the faulting page as read-heavy or write-heavy using the dual bCBF (\S\ref{subsubsec:rw-class}), and checks the current CXL traffic pattern. The admission control policy is summarized in Table~\ref{tab:admission-policy}: \name migrates pages whose type matches the dominant traffic, and holds back the rest. The admission control runs before the splitting stage: the eBPF hook decides whether to admit the page for migration and only admitted pages proceed to the splitting pipeline.

%% file: text/evaluation.tex
\section{Evaluation}
\label{sec:eva}
\textcolor{checked}{\name comprises $\sim$6,800 lines of user-space code (eBPF programs, loaders, daemons, and analyzers) and 540 lines of kernel modifications to Linux v6.12 for eBPF hooks and page splitting. Since neither TPP nor Colloid supports mTHP, we re-implemented both on Linux v6.12, requiring $\sim$270 and 1,160 lines of kernel changes, respectively.}  

\subsection{Evaluation Methodology}
\label{sec:eval_method}
\textbf{Evaluation platforms.}
We use two servers.

\begin{denseitemize}
    \item A server with CXL memory expansion with dual-socket AMD EPYC 9555s (64 cores per socket). Each socket has 8$\times$32\,GB DDR5 DRAM as local memory. Socket~0 connects to a CXL Type-3 Memory Expansion device (128\,GB DDR4) over PCIe~5.0, providing a full-duplex link. 
    
    \item A server with Intel Optane PMEM with dual-socket Intel Xeon Gold 6252 (24 cores per socket). Each socket has 6$\times$16\,GB DDR4 DRAM as the fast tier and 6$\times$128\,GB Intel Optane PMEM as the slow tier. 
\end{denseitemize}

By default, we use the CXL server. To isolate the impact of page migration from cross-socket NUMA effects, we use Socket~0 only, following efforts in~\cite{lee2023memtis, maruf2023tpp, atc24_hm}. The hardware event sampling rate is 4{,}000 samples/sec per CPU.

\textbf{System integration.} To evaluate the portability of \name, we integrate \name into three kernel-level solutions: AutoNUMA, TPP, and Colloid. \textcolor{checked}{For AutoNUMA, we use the vanilla Linux kernel v6.12. For TPP, we change AutoNUMA's hotness detection from NUMA hint fault latency to an LRU active-list check and add TPP's asynchronous background demotion mechanism~\cite{maruf2023tpp}. For Colloid, we use its TPP-based implementation~\cite{vuppalapati2024colloid} and replace Intel CHA-based latency monitoring with AMD Data Fabric performance counters.} 

\textbf{Baselines.} On the CXL server, we integrate \name into AutoNUMA, TPP, and Colloid, and compare against 
three baselines: 2MB-alloc (2\,MB THP allocation), 4KB-alloc (4\,KB base-page allocation), and 4KB-split.
On the PMEM-enabled server, we additionally compare \name against MEMTIS~\cite{lee2023memtis}, the state-of-the-art system for huge-page-aware page migration. 
We cannot evaluate MEMTIS on the CXL platform because MEMTIS relies on Intel PEBS events to distinguish memory access sources (i.e., local DRAM or remote DRAM) at the hardware level.  AMD IBS, used on our CXL server, does not provide equivalent hardware performance counters. Hence, we use 4KB-split to approximate MEMTIS on the CXL platform by forcing \name to split a THP into 4\,KB pages upon the same skew detection.

\input table/benchmarks

\textbf{Workloads.} We evaluate \name on 17 memory-intensive benchmarks shown in Table~\ref{tab:benchmarks}, spanning three categories: 12 graph-processing workloads (from GAP~\cite{gap}), an in-memory database engine (Silo-YCSB~\cite{silo}), and 4 HPC benchmarks (from NPB~\cite{nas}).
All are used extensively in related work~\cite{10.1145/3712285.3759816,lee2023memtis,raybuck2021hemem,maruf2023tpp,eurosys24:mtm,osdi25_aol,vuppalapati2024colloid}.
GAP and NPB benchmarks run with 12 threads; Silo-YCSB runs with 8 threads since it crashes with 12.
Each benchmark runs in a dedicated cgroup with controlled CPU and memory limits, and its fast memory is capped at 50\% of its memory footprint.

\textbf{Memory contention.} To emulate the bandwidth pressure commonly observed in virtualized environments, we run a memory stressor on the slow tier alongside benchmarks. The stressor spawns antagonist threads that issue sequential, non‑temporal memory accesses using \texttt{movntdqa}/\texttt{movntdq} instructions over a buffer allocated and pinned in the slow tier. By bypassing the CPU cache hierarchy, these accesses maximize memory bus utilization. The antagonist threads are pinned to dedicated CPU cores separate from the benchmark to avoid direct CPU contention. We vary the number of antagonist threads to generate sustained background traffic of approximately 0, 5, 10, and 20\,GB/s. Unless otherwise noted, all experiments run under 20\,GB/s of background traffic; \S\ref{sec:eval_contention} evaluates \name across various contention levels.

\subsection{Evaluation of THP Splitting}
\label{sec:eval_cxl_splitting}

\begin{figure}[!t]
  \centering
\includegraphics[width=1\columnwidth]{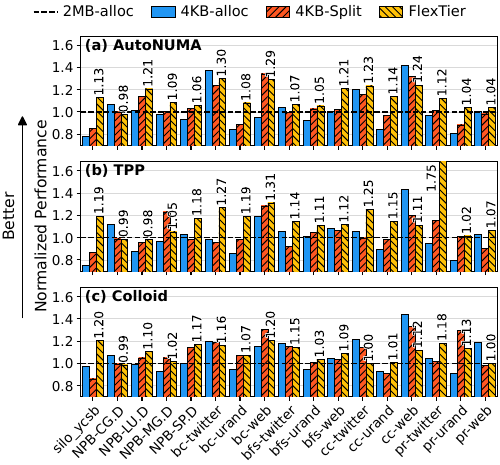}
\vspace{-20pt}
  \caption{Performance of \name and of the baselines on the CXL platform, for three tiering systems. All bars are normalized to 2MB-alloc (dashed line).}
  \label{fig:cxl_splitting}
  \vspace{-15pt}
\end{figure}


\subsubsection{Comparison Against Each Baseline} 
\name outperforms 2MB-alloc by 13.0\% on AutoNUMA, 15.9\% on TPP, and 9.3\% on Colloid in geomean (Figure~\ref{fig:cxl_splitting}). Graph workloads and Silo-YCSB benefit the most from \name. For example, bc-twitter improves by 30\%, bc-web by 29\%, cc-web by 24\%, cc-twitter by 23\%, and bfs-web by 21\%. Silo-YCSB improves consistently across all three systems (13\% on AutoNUMA, 19\% on TPP, and 20\% on Colloid). These workloads exhibit highly skewed intra-THP access: for example, the power-law degree distributions of the Twitter and Web graphs concentrate accesses on a small set of high-degree vertices.  
By splitting THPs and migrating only the hot subfolios, \name avoids wasting scarce fast-tier capacity on cold data. 
\name does not improve CG.D. 
CG's accesses are highly skewed \emph{across} THPs but nearly uniform \emph{within} each THP (\S\ref{sec:eval_fidelity}), providing little chances to split pages.

Compared to 4KB-split, \name improves by 8.4\% on AutoNUMA, 13.0\% on TPP, and 0.9\% on Colloid in geomean.  Splitting all the way to base pages forfeits the page-fault and TLB benefits of huge pages entirely, even when an intermediate size would have been sufficient to capture the hot region.  However, both \name and 4KB-split outperform 2MB-alloc on the most skewed workloads. On AutoNUMA, for example, bc-web reaches $1.29\times$ under \name and $1.34\times$ under 4KB-split, and cc-web reaches $1.24\times$ and $1.32\times$, because of concentrated hot subpages within THPs.

\name also outperforms 4KB-alloc by 13.4\% on AutoNUMA, 16.6\% on TPP, and 3.1\% on Colloid in geomean. Disabling huge pages makes each migration move only one 4\,KB page, but it charges the costs of base-page translation and page-faults to the entire address space. 

\subsubsection{Analysis by Memory Tiering Software} 
TPP benefits the most among the three memory tiering systems, with a 15.9\% geomean improvement. Unlike AutoNUMA, which infers page hotness from hint fault latency (i.e., the time between unmapping a page and the subsequent access fault), TPP relies on an LRU-based heuristic: a page is deemed hot if it resides on the kernel’s active LRU list~\cite{maruf2023tpp}. However, for a 2\,MB THP comprising 512 subpages, a single access to any subpage causes the entire folio to be referenced and promoted to the active list. Consequently, TPP’s LRU check is rendered ineffective at THP granularity—nearly all THPs are set to be promoted regardless of how little of the folio is actually accessed. This behavior leads TPP to over-promote cold data into the fast tier, populating it with folios that are only sparsely utilized. Splitting restores the discriminative power of the LRU filter. A 64\,KB subfolio, for example, contains only 16 subpages, allowing cold subfolios to age into the inactive list and be correctly classified as cold. As a result, TPP experiences the largest performance gains. 


Colloid shows the smallest improvement (9.3\% geomean) because its access-latency–balanced promotion policy selects pages more judiciously. Nevertheless, \name still yields gains: silo\_ycsb improves by 20\%, bc-web by 20\%, and pr-twitter by 18\%. These results highlight that even a well-tuned page selection policy cannot avoid bandwidth waste when migration is constrained to coarse-grained 2\,MB pages. 

\input{table/table_split_distribution_selected}

\subsubsection{How Does \name Split Pages?} 


See Table~\ref{tab:split_dist_selected}. First, no single mTHP size fits all workloads. For instance, bc-twitter distributes its splits across four mTHP sizes (64\,KB -- 512\,KB), reflecting that the THP's hot region extent varies across execution phases. This underscores that a static, one-size-fits-all splitting granularity would leave substantial performance untapped; the dynamic size selection is essential. 

Second, hot subfolios are typically a minority. Across the workloads, the fraction of hot subfolios ranges from just 9.1\% (cc-twitter) to 53.2\% (bfs-web). Without splitting, the kernel would promote entire 2\,MB THPs, wasting up to 91\% of migration bandwidth and fast-tier capacity on cold data.

\subsection{Why does Splitting Help? Further Analysis from the Perspective of Page Migration}
\label{sec:eval_deep}

\input{table/table_split_vs_nosplit}

We analyze page migration statistics collected from kernel vmstat counters on AutoNUMA. See Table~\ref{tab:split_vs_nosplit}. 


The most pronounced change is the reduction in migration failures. Without splitting, the kernel attempts to migrate entire 2\,MB THPs; when the destination node cannot provide a contiguous 2\,MB free region, the migration fails. For bc-twitter, 18 million migration attempts fail without splitting, whereas with splitting only 80 thousand fail—a 224$\times$ reduction. Similar trends hold across all workloads. 


The underlying reason is clear. Splitting a 2\,MB THP into, for example, 64\,KB mTHP subfolios reduces migration’s allocation requirement to just 64\,KB of contiguous free space on the destination node—an order-of-magnitude easier constraint to satisfy. As a result, far fewer migrations fail, allowing AutoNUMA to effectively relocate hot data.


This sharp reduction in failures translates directly into more successful migrations. For bc-twitter, the number of successful migrations nearly doubles, increasing from 1.26\,M to 2.44\,M (1.9$\times$). For cc-web, successful migrations grow from 1.94\,M to 4.28\,M (2.2$\times$), and for cc-twitter from 1.24\,M to 3.80\,M (3.1$\times$). As a result, a larger fraction of hot data is successfully placed in the fast tier, while cold subpages remain in the slow tier. Across all 12 workloads, splitting increases the number of successful migrations by 1.0$\times$–3.1$\times$ while reducing migration failures by up to 777$\times$.


In summary, splitting improves performance through two reinforcing effects: it lowers the bar for migration success by reducing allocation size requirements, and it improves migration precision by ensuring that only hot subfolios are moved. Together, these effects enable AutoNUMA to migrate hot data more reliably and more selectively. 

\input{table/table_tlb_hintfault}

\subsection{How does \name Retain Huge-Page Benefits? }
\label{sec:eval_tlb_fault} 

See Table~\ref{tab:tlb_hintfault}. For L2 dTLB misses, 4KB-alloc incurs $30.4\times$ that of 2MB-alloc in geomean; 4KB-split incurs $3.3\times$, 
because it splits only the THPs that a migration decision touches and leaves the rest 2\,MB-mapped. \name stays closest to 2MB-alloc. 
We observe a similar trend for L1 dTLB MPKI.

For NUMA hint faults, 4KB-alloc generates $166.9\times$ of 2MB-alloc in geomean. 4KB-split generates $20.3\times$ of 2MB-alloc, because a THP split into base pages leaves 512 separately-faulted folios to the kernel. \name generates only $3.0\times$ of 2MB-alloc---$1/7$ of 4KB-split's faults.
These results show that \name makes finer migration decisions while retaining huge-page benefits, in TLB misses and NUMA hint faults.

\subsection{Fidelity of the Global Histogram}
\label{sec:eval_fidelity}
We compare the split decisions made with a global histogram against those made by the per-THP histogram references. 
The per-THP histogram uses the same eBPF-based kernel profiling as \name (\S\ref{subsubsec:ebpf-profiling}) but builds a 512-bin access histogram per THP. The per-THP histogram can be based on hardware sampling or PTE scanning.

The PTE scanning approach must disable THP so that accessed bits are scanned at base-page granularity. It builds a 512-bin histogram per THP. Note that it cannot be implemented as a runtime system: with THP enabled, a PMD-mapped 2\,MB THP is covered by one PMD entry with a single accessed bit, leaving no per-subpage PTEs to scan. Hence, 
we log every 2\,MB THP migration event triggered by NUMA faults (i.e., decision-relevant) and replay the events offline. 
We use three types of memory intensive workloads, Zipfian key--value (silo\_ycsb), skewed graph (bc-twitter), and HPC (NPB-CG.D).

\input{table/fidelity}

Table~\ref{tab:fidelity} shows the results. First, \textbf{folding in the global histogram is near-lossless, compared with the per-THP histogram.} On silo\_ycsb, with Zipfian accesses, the global histogram and per-THP histogram make the same decision---splitting all THPs into 64\,KB pages. The PTE-scan histogram confirms this decision, splitting 74.8\% of THPs into 64\,KB pages. The global histogram and the per-THP histogram outperform the 2\,MB baseline by 13\% and 11\%, respectively. On bc-twitter, 92\% of the global histogram's decisions either match the per-THP histogram's or are at most one size-class coarser: the per-THP histogram splits 92.1\% of faulted THPs into 64\,KB pages, while the global histogram splits 67.3\% into 64\,KB and 25.5\% into 128\,KB. Both outperform the 2\,MB baseline, by 30\% and 19\%, respectively. Folding in the global histogram can lead to a coarser split choice, but the coarser split does not degrade performance: the global histogram outperforms the per-THP histogram by 9.1\%, because the retained larger mappings preserve large-page benefits without admitting cold subpages. The PTE-scan histogram treats most THPs in bc-twitter as uniformly accessed because an accessed bit records only \emph{whether} a page was touched, not \emph{how often}: accessed subpages appear equally ``touched'' in a scan window regardless of how frequently they are accessed.

Second, \textbf{the hardware-sampling-based per-THP histogram fails to recognize uniform intra-THP access} due to the sampling sparsity ( \S\ref{subsec:profiling}). As a result, most THPs are split into the smallest page size on all workloads. On NPB-CG.D, while the global histogram and the PTE-scan histogram migrate 2\,MB THPs without splitting, the per-THP histogram splits 81.7\% of them into 64\,KB pages, reducing throughput by 13\% (0.87$\times$); the global histogram matches the baseline (0.98$\times$). We further find that CG.D's accesses are highly skewed across THPs but flat within each THP; the per-THP histogram misses the intra-THP flatness due to its sample sparsity.

\subsection{Evaluation with Memory Bandwidth Contention}
\label{sec:eval_contention}

\begin{figure}[t]
  \centering
  \includegraphics[width=\columnwidth]{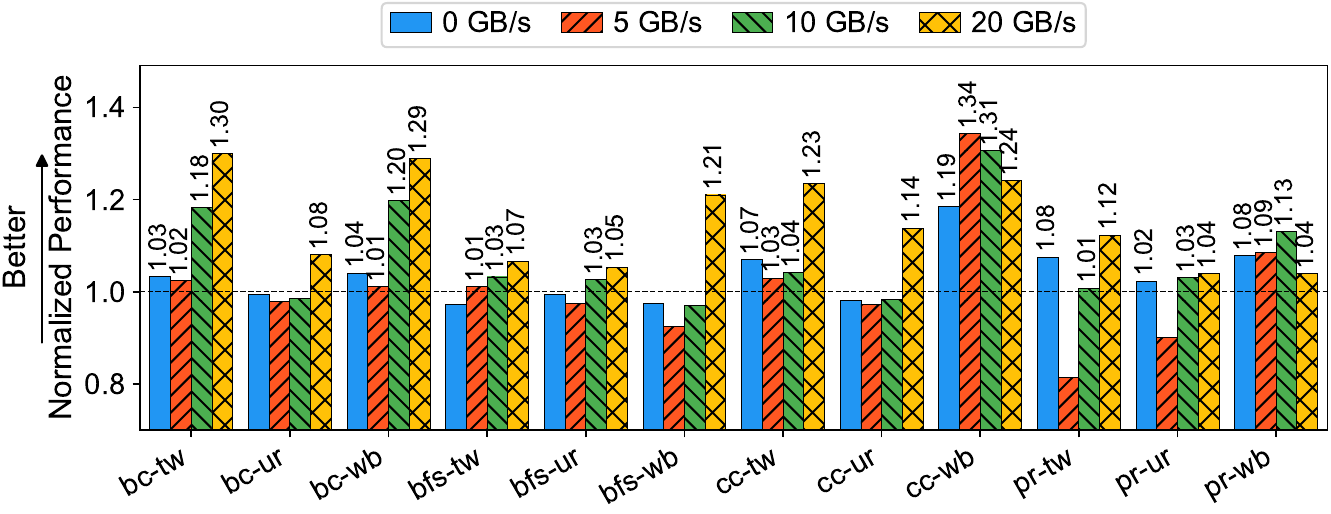}
  \caption{Effects of CXL memory bandwidth contention on splitting benefit (AutoNUMA). The performance is normalized to 2MB-alloc without THP splitting.}
  \label{fig:contention}
\vspace{-20pt}

\end{figure}

\name activates THP splitting only when slow-tier bandwidth becomes contended (\S\ref{subsec:splitting}). To validate this design choice and motivate the contention level used in our main evaluation (\S\ref{sec:eval_cxl_splitting}), we run the 12 GAP workloads on AutoNUMA under four levels of CXL memory bandwidth pressure: 0, 5, 10, and 20\,GB/s of sustained background traffic, generated by a memory-streaming process on the CXL device.

See Figure~\ref{fig:contention}. 
Splitting improves throughput only when slow-tier bandwidth is scarce: compared to 2MB-alloc, the geomean improvement is 
$-$0.1\% at 5\,GB/s (6 of the 12 workloads), 7.0\% at 10\,GB/s (9 of the 12 workloads), and 14.7\% at 20\,GB/s (all 12 workloads). 

When bandwidth is abundant, migrating cold subpages alongside hot ones within a 2\,MB THP does not displace useful traffic, so the added TLB misses, page faults and tracking pressure outweighs the savings in migration bandwidth. Under contention, migrating a 2\,MB page which is only 20\% hot wastes 1.6\,MB of scarce bandwidth, and the savings outweigh the TLB and page fault penalty. Hence we use the contention level of 20\,GB/s in \S\ref{sec:eval_cxl_splitting}, as it reflects a realistic scenario in which multiple memory‑intensive workloads share a CXL device and bandwidth becomes bottleneck. These results validate \name's contention‑aware 
policy: \name preserves 2\,MB THPs under low contention, and enables splitting only when bandwidth savings outweigh the associated TLB costs.

\begin{figure}[!t]
  \centering
  \includegraphics[width=\columnwidth]{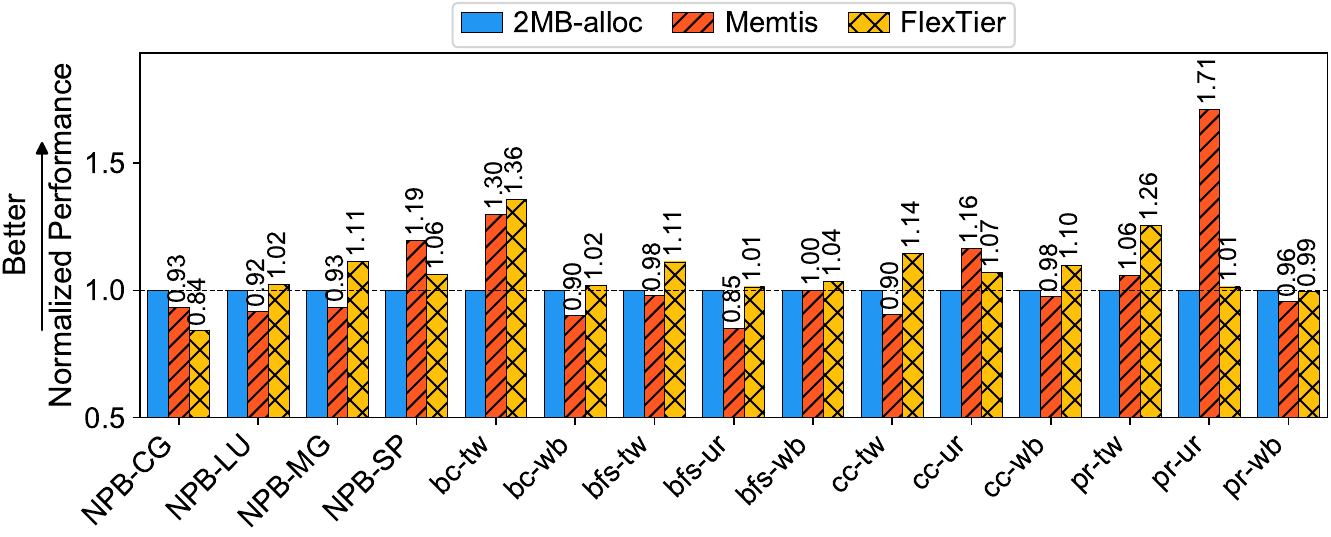}
  \caption{Performance of \name/MEMTIS on the PMEM platform, normalized to that of 2MB-alloc on AutoNUMA.} 
  \label{fig:intel_splitting}
\vspace{-20pt}

\end{figure}

\subsection{Evaluation on PMEM}
\label{sec:eval_optane}



To demonstrate that \name's design generalizes beyond CXL, 
we compare \name with 2MB-alloc and the real MEMTIS (rather than the 4KB-split approximation used on the CXL platform).  
\textcolor{checked}{We enable only the THP splitting policy, because maintaining mixed read/write traffic does not benefit performance given PMEM's half-duplex DDR interface.}


See Figure~\ref{fig:intel_splitting}. 
\name outperforms the 2MB-alloc baseline by 7.1\% in geomean, 
with the largest gains on the graph workloads with highly skewed access patterns (e.g., bc-twitter by 36\% and pr-twitter by 26\%).
\name outperforms MEMTIS on 11 of 15 workloads; this improvement is most evident on NPB workloads, which traverse large arrays and are particularly sensitive to page fault, page tracking, and TLB overheads. The performance trends on PMEM closely mirror those on CXL, confirming that THP splitting addresses a fundamental limitation of THP-based migration---namely, the mismatch between the coarse-grained migration and fine-grained, sub-page access skew.

\subsection{Evaluation with Traffic Balancing}
\label{sec:rw}

We evaluate \emph{BalanceTraffic}, the duplex-aware migration admission policy in  \S\ref{subsec:admission}, under two regimes: high contention at 20\,GB/s and moderate contention at 10\,GB/s.

\textbf{High contention (20\,GB/s).}
See Figure~\ref{fig:rebalance}(a). 
BalanceTraffic does not complement splitting under high contention. On AutoNUMA, enabling Split alone improves performance by 13.0\%, while Split+BalanceTraffic reduces it to 6.6\%. The effect is more pronounced on TPP: Split alone achieves 15.9\%, 
whereas Split+BalanceTraffic yields only 1.2\%.

This is due to the following: Under 20\,GB/s, 
aggregate bandwidth is already the bottleneck, and performance is determined by whether hot pages are placed in fast tier. 
Rebalancing read/write traffic across CXL’s full‑duplex channels provides little additional benefit---the limiting factor is total bandwidth, not directional imbalance. Moreover, BalanceTraffic’s selective admission policy can be counterproductive: by deferring migrations of hot pages whose R/W classification does not match the dominant traffic direction (Table~\ref{tab:admission-policy}), it prevents them from reaching fast tier, where they are  needed.


\textcolor{checked}{When applied alone (without Split), BalanceTraffic yields a modest improvement. On TPP in particular, only 7 workloads see improvements while 10 workloads degrade, and the overall geomean gain is only 2.5\%.}

\textbf{Moderate contention (10\,GB/s).}
The results improve: BalanceTraffic achieves a geomean improvement of 5.2\% on AutoNUMA and 4.7\% on TPP, with 13 of 17 workloads improved (Figure~\ref{fig:rebalance}(b)). 
With more bandwidth headroom, the full-duplex rebalancing has room to take effect without starving hot pages of fast-tier placement. As such, selectively deferring promotions that would add traffic to the non-bottleneck CXL channel shows benefits.

%% file: table/benchmarks.tex
\begin{table}[t]
\caption{Benchmarks used for evaluation.}
\label{tab:benchmarks}
\centering
\resizebox{\columnwidth}{!}{
\begin{tabular}{lllr}
\toprule
\textbf{Suite} & \textbf{Benchmark} & \textbf{Input / Configuration} & \textbf{Footprint}  \\
\midrule
\multirow{4}{*}{GAP~\cite{gap}}
  & BC  & \multirow{4}{*}{Twitter~/ Uniform Random~/ Web} & \multirow{4}{*}{12--24\,GB} \\
  & BFS &  &  \\
  & CC  &  &  \\
  & PR  &  &  \\
\midrule
\multirow{6}{*}{NPB~\cite{nas}}
  & LU.D  & Lower-Upper Gauss-Seidel solver   & 58\,GB \\
  & SP.D  & Scalar Penta-diagonal solver       & 76\,GB \\
  & CG.D  & Conjugate Gradient                 & 16\,GB \\
  & MG.D  & Multigrid                          & 28\,GB \\
\midrule
Silo~\cite{silo}
  & YCSB  & YCSB, scale factor 200K            & 29\,GB \\

\bottomrule
\end{tabular}
}
\vspace{-1.5em}
\end{table}

%% file: table/table_split_distribution_selected.tex
\begin{table}[t]
\centering
\caption{Page splitting for four workloads with the largest improvement on AutoNUMA.
\emph{Split target} shows the number of split events for each mTHP size;
\emph{Hot Subfolios} is the fraction of resulting subfolios classified as hot.}
\label{tab:split_dist_selected}
\resizebox{\columnwidth}{!}{
\footnotesize
\begin{tabular}{l|rrrrr|r}
\toprule
 & \multicolumn{5}{c|}{Split target (count) } & \\
Workload & 64KB & 128KB & 256KB & 512KB & 1MB & Hot Subfolios (\%)  \\
\midrule
cc-twitter  & 4,101 & 21    & 0     & 0     & 0     & 9.1\%  \\
cc-web      & 0     & 6,013 & 147   & 1,470 & 0     & 21.9\% \\
bc-twitter  & 1,998 & 186   & 1,907 & 1,103 & 0     & 22.5\% \\
bfs-web     & 0     & 0     & 422   & 1,170 & 881   & 53.2\% \\
\bottomrule
\end{tabular}
}
\vspace{-20pt}
\end{table}

%% file: table/table_split_vs_nosplit.tex
\begin{table}[t]
\centering
\caption{Page migration statistics: NoSplit  vs. THP Split.}
  \vspace{-1em}
\label{tab:split_vs_nosplit}
\resizebox{\columnwidth}{!}{
\footnotesize
\begin{tabular}{l|rr|rr|rr}
\toprule
 & \multicolumn{2}{c|}{Page Demotion} & \multicolumn{2}{c|}{Page Promotion Fail} & \multicolumn{2}{c}{Page Promotion Success} \\
Workload & NoSplit & THP Split & NoSplit & THP Split & NoSplit & THP Split \\
\midrule
bc-twitter  & 1.10\,M & 1.34\,M & 18.01\,M &  0.08\,M & 1.26\,M & 2.44\,M \\
bc-urand    & 1.61\,M & 1.96\,M &  8.46\,M &  0.01\,M & 2.32\,M & 3.34\,M \\
bc-web      & 1.44\,M & 2.20\,M &  6.58\,M &  1.18\,M & 1.94\,M & 3.79\,M \\
bfs-twitter & 1.48\,M & 1.07\,M &  4.53\,M &  2.18\,M & 1.63\,M & 1.71\,M \\
bfs-urand   & 1.81\,M & 2.02\,M &  7.03\,M &  1.60\,M & 2.19\,M & 3.25\,M \\
bfs-web     & 1.54\,M & 1.82\,M &  2.66\,M &  0.08\,M & 2.09\,M & 2.77\,M \\
cc-twitter  & 1.07\,M & 2.20\,M &  7.66\,M &  0.69\,M & 1.24\,M & 3.80\,M \\
cc-urand    & 1.81\,M & 2.19\,M &  1.28\,M &  0.004\,M & 2.31\,M & 2.91\,M \\
cc-web      & 1.56\,M & 2.54\,M &  6.00\,M &  0.44\,M & 1.94\,M & 4.28\,M \\
pr-twitter  & 1.07\,M & 1.12\,M &  5.69\,M &  2.95\,M & 1.37\,M & 1.80\,M \\
pr-urand    & 1.71\,M & 1.71\,M &  0.53\,M &  0.19\,M & 2.08\,M & 2.17\,M \\
pr-web      & 1.49\,M & 1.49\,M &  1.58\,M &  1.23\,M & 2.02\,M & 1.92\,M \\
\bottomrule
\end{tabular}
}
\vspace{-20pt}
\end{table}

%% file: table/table_tlb_hintfault.tex
\begin{table}[t]
\centering
\caption{L2 dTLB misses per 1K instructions (MPKI) and NUMA hint faults on AutoNUMA, for the 12 GAP workloads.}
  \vspace{-1em}
\label{tab:tlb_hintfault}
\resizebox{\columnwidth}{!}{
\footnotesize
\begin{tabular}{l|rrrr|rrrr}
\toprule
 & \multicolumn{4}{c|}{L2 dTLB MPKI} & \multicolumn{4}{c}{NUMA hint faults} \\
 & 4KB & 2MB & 4KB & \multirow{2}{*}{\name} & 4KB & 2MB & 4KB & \multirow{2}{*}{\name} \\
Workload & alloc & alloc & split & & alloc & alloc & split & \\
\midrule
bc-twitter   &   21.18 &    0.22 &    1.50 &    1.37 & 10,228\,K & 67\,K & 7,639\,K & 332\,K \\
bc-urand     &   21.56 &    0.76 &    1.68 &    1.82 & 9,744\,K & 28\,K & 3,579\,K & 390\,K \\
bc-web       &    1.32 &    0.07 &    0.50 &    0.32 & 5,403\,K & 16\,K & 1,060\,K & 104\,K \\
bfs-twitter  &    2.30 &    0.14 &    0.50 &    0.25 & 2,412\,K & 15\,K & 1,216\,K & 44\,K \\
bfs-urand    &    3.06 &    0.18 &    0.43 &    0.30 & 7,724\,K & 34\,K & 683\,K & 73\,K \\
bfs-web      &    0.29 &    0.05 &    0.14 &    0.09 & 3,550\,K & 14\,K & 620\,K & 38\,K \\
cc-twitter   &    5.23 &    0.11 &    0.52 &    0.30 & 3,297\,K & 29\,K & 3,043\,K & 410\,K \\
cc-urand     &    4.59 &    0.12 &    0.39 &    0.21 & 5,007\,K & 45\,K & 667\,K & 41\,K \\
cc-web       &    0.45 &    0.07 &    0.26 &    0.15 & 3,267\,K & 31\,K & 1,301\,K & 95\,K \\
pr-twitter   &   65.66 &    0.15 &    0.30 &    0.23 & 6,327\,K & 75\,K & 46\,K & 117\,K \\
pr-urand     &   56.93 &    0.26 &    0.56 &    0.38 & 12,261\,K & 59\,K & 73\,K & 49\,K \\
pr-web       &    0.33 &    0.05 &    0.13 &    0.08 & 1,544\,K & 13\,K & 19\,K & 15\,K \\
\bottomrule
\end{tabular}
}
\vspace{-10pt}
\end{table}

%% file: table/fidelity.tex
\begin{table}[t]
\centering
\caption{Split-size decisions (\% of decision events) of the PTE-scan hist, the global histogram, and the per-THP hist. Norm.\ Perf.\ is workload throughput normalized to the no-split 2\,MB THP baseline; the PTE-scan hist\ is offline-only and thus has no throughput.}
\vspace{-10pt}
\label{tab:fidelity}
\resizebox{\columnwidth}{!}{
\footnotesize
\begin{tabular}{ll|rrrrrr|r}
\toprule
Workload & Profiling & No split & 1\,MB & 512\,KB & 256\,KB & 128\,KB & 64\,KB & Norm.\ Perf. \\
\midrule
 \multirow{3}{*}{\shortstack[l]{silo\_ycsb}}
 & PTE-scan hist    & 0.9 & 0.6 & 2.4 & 6.8 & 14.4 & \textbf{74.8} & --- \\
 & Global histogram & 0.0 & 0.0 & 0.0 & 0.0 & 0.0 & \textbf{100.0} & 1.13 \\
 & Per-THP hist     & 0.0 & 0.0 & 0.0 & 0.0 & 0.0 & 100.0 & 1.11 \\
\midrule
\multirow{3}{*}{\shortstack[l]{bc-twitter}}

 & PTE-scan hist & 81.6 & 0.4 & 0.9 & 3.3 & 7.9 & 5.8 & --- \\
 & Global histogram & 1.1 & 0.0 & 0.0 & 6.0 & 25.5 & 67.3 & 1.30 \\
 & Per-THP hist     & 0.1 & 0.1 & 0.3 & 0.6 & 6.8 & 92.1 & 1.19 \\
\midrule
\multirow{3}{*}{\shortstack[l]{NPB-CG.D}}
 & PTE-scan hist    & \textbf{94.5} & 0.3 & 0.3 & 0.0 & 0.0 & 4.8 & --- \\
 & Global histogram & \textbf{100.0} & 0.0 & 0.0 & 0.0 & 0.0 & 0.0 & 0.98 \\
 & Per-THP hist     & 0.8 & 0.0 & 2.4 & 7.2 & 7.9 & 81.7 & 0.87 \\
\bottomrule
\end{tabular}
}
\vspace{-10pt}
\end{table}

%% file: text/related_work.tex
\section{Related Work} 
\label{sec:related}

\begin{figure}[!t]
  \centering
  \includegraphics[width=1\columnwidth]{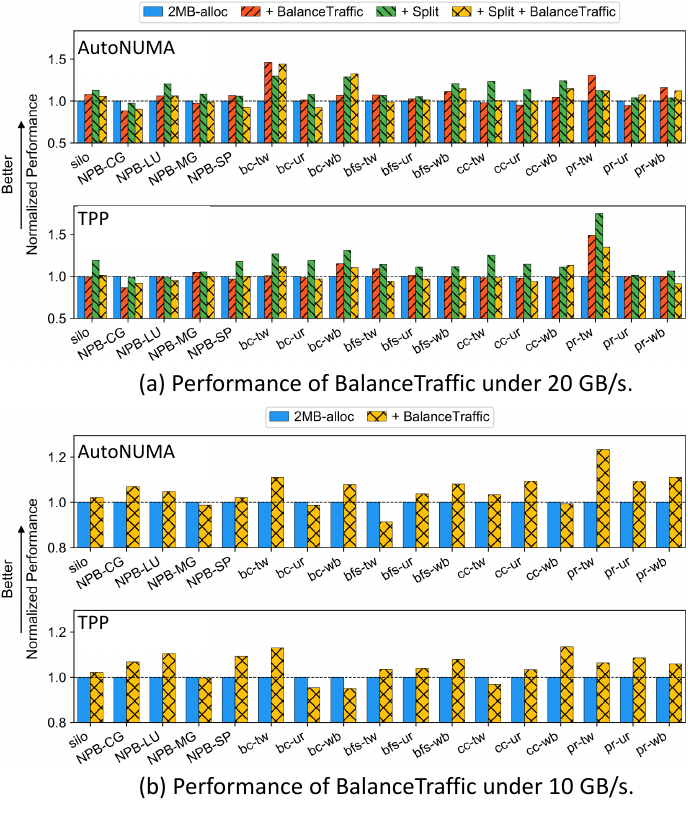}
  \caption{Performance of BalanceTraffic, normalized to the AutoNUMA and TPP 2MB-alloc baselines.}
  \label{fig:rebalance}
\end{figure}

\textbf{Memory tiering systems.} 
AutoNUMA~\cite{autonuma} migrates pages by access recency, and TPP~\cite{maruf2023tpp} extends it with proactive demotion and LRU active-list-based promotion. 
HeMem~\cite{raybuck2021hemem}, MEMTIS~\cite{lee2023memtis}, MTM~\cite{eurosys24:mtm}, and FlexMem~\cite{atc24_hm} migrate by access frequency. Memstrata~\cite{10.5555/3691938.3691941} utilizes Intel Flat Memory mode for cache-level placement, and Nomad~\cite{xiang2024nomad} builds non-exclusive memory tiering. Colloid~\cite{vuppalapati2024colloid} balances access latency across tiers. HybridTier~\cite{10.1145/3676642.3736119} combines long-term access frequency and short-term access momentum, while M5~\cite{m5_asplos2025} tracks hot data in the CXL controller. 
Alto~\cite{osdi25_aol} regulates page promotion using a metric that accounts for memory-level parallelism. The above efforts decide \emph{which} pages to migrate and \emph{when}, at a fixed page granularity. In contrast, \name decides \emph{at what granularity} to migrate and \emph{in which direction} to spend link bandwidth, and makes both decisions programmable. \name can be integrated into other solutions (e.g., AutoNUMA, TPP, and Colloid as we did). 

\textbf{CXL memory.} 
{Prior work uses CXL for point-to-point and collective communication~\cite{10.1145/3712285.3759816,xu2026ccclnodespanninggpucollectives}, characterizes real CXL hardware~\cite{ipdps25:cxl,10.1145/3676641.3715987,eurosys24_cxl,mao2024cxlinterferenceanalysischaracterizationmodern,micro23_cxl}, offloads AI tensors~\cite{11408577,yang2025beluga}, and backs databases~\cite{10.1145/3722212.3724460}. CXL also supports memory pooling and cross-host sharing~\cite{10.1145/3575693.3578835,10.1145/3779212.3790149,10.1145/3779212.3790150,10.1145/3779212.3790121,10.1145/3779212.3790245}, function-as-a-service~\cite{patil2023apta}, RPC data transmission~\cite{ma2024hydrarpc}, reliable and secure memory~\cite{11408527}, and replication for partial-failure tolerance~\cite{zhu2024lupin}. \name provides  optimizations on CXL memory.}

\textbf{eBPF for memory optimization.} 
eBPF has been used to customize memory management policies, such as huge page placement, page fault handling, and page table designs~\cite{mores2024ebpf,zussman2024custom}, as well as page cache customization, prefetching, and swapping policies~\cite{lee2026p2cache,zussman2025cache_ext,yelam2025pageflex,cao2024fetchbpf}. 
CXLAimPod~\cite{yang2025cxlaimpodcxlmemoryneed} exploits the full-duplex CXL link, but balances the bidirectional stream through \emph{thread} placement rather than \emph{data} placement as in \name: CXLAimPod's \texttt{sched\_ext}-based CPU scheduler co-places read- and write-heavy threads on the same CPU cluster, while all memory stays bound to a single CXL node and page migration across tiers is explicitly out of scope.
In contrast, \name leverages eBPF to provide policy flexibility for page migration across memory tiers.


%% file: text/conclusion.tex
\section{Conclusion}
Memory tiering architectures are rapidly diversifying, which calls for consideration of page-migration interactions with other system components and memory architecture details. 
This paper discusses how to improve the memory tiering system to accommodate emerging changes in the memory architecture, especially from the perspectives of page size and memory access distribution. \name enables sufficient flexibility to develop novel policies in diverse environments, such as more than two memory tiers or CXL-SSD, which could be implemented in the future. 
